\documentclass[aps,twocolumn,prb,superscriptaddress,sloatfix]{revtex4-2}
\usepackage{amssymb,bbm,bm}
\usepackage{graphicx}
\usepackage{bm}
\usepackage{amsmath}
\usepackage[usenames]{color}
\usepackage{epsfig}
\usepackage{hyperref}
\usepackage{subfigure}
\usepackage[dvipsnames]{xcolor}
\usepackage[sans]{dsfont}
\hypersetup{colorlinks=true, citecolor=blue,
linkcolor=blue,urlcolor=blue }

\usepackage{amsmath,amsfonts,amssymb,amsthm,bm}
\usepackage{float}
\usepackage{hyperref}
\usepackage{graphicx}
\usepackage{subfigure}

\begin{document}

\title{AC conductivity and magnetic dichroism of two-dimensional antiferromagnetic Dirac semimetals}

\author{Reza Sepehrinia}\email{sepehrinia@ut.ac.ir}
\affiliation{Department of Physics, University of Tehran, Tehran 14395-547, Iran}

\author{Siavash Eskandari}
\affiliation{Department of Physics, University of Tehran, Tehran 14395-547, Iran}

\author{Alireza Qaiumzadeh}
\affiliation{Center for Quantum Spintronics, Department of Physics, Norwegian University of Science and Technology, NO-7491 Trondheim, Norway}

\begin{abstract}
We investigate the magneto-optical properties of two-dimensional nonsymmorphic Dirac semimetals in the presence of antiferromagnetic order. Using the Kubo formula, we calculate the conductivity tensor of two-dimensional CuMnAs, a prototype antiferromagnetic Dirac material, as a function of light frequency. From the finite-frequency conductivity tensor, we derive the dynamic dielectric function and magnetic linear dichroism, demonstrating how they are influenced by the orientation of the N{\'e}el order parameter. Adjusting the N{\'e}el vector changes both the sign and amplitude of the system's magneto-optical response. We propose that magnetic linear dichroism spectroscopy is a powerful technique for determining the orientation of the N{\'e}el vector.
\end{abstract}
\maketitle

\section{Introduction}
On-demand manipulation and readout of the staggered  N{\'e}el order parameter in antiferromagnetic (AFM) systems are challenging problems in the emerging field of AFM-based spintronics \cite{jungwirth2016antiferromagnetic,baltz2018antiferromagnetic}.
Both theoretically and experimentally, spin-orbit and spin-transfer torques have been demonstrated to be finite in AFM systems under certain symmetry conditions \cite{PhysRevB.104.224414,PhysRevLett.110.127208,STT,vsmejkal2017electric,vsmejkal2018topological,manchon2019current,PhysRevB.81.144427,PhysRevLett.117.017202,wadley2016electrical}.
It was shown that in AFM systems with preserved combined $\mathcal{PT}$ symmetry, while the time-reversal $\mathcal{T}$ symmetry and the inversion $\mathcal{P}$ symmetry are
each broken, staggered spin-orbit torques acting on the magnetic sublattices may reorient the AFM N{\'e}el vector \cite{wadley2016electrical,vsmejkal2017electric,STT,PhysRevLett.117.017202,PhysRevB.104.224414,PhysRevLett.113.157201,cumnasDirac}.

Two-dimensional (2D) Dirac materials are a recently discovered class of 2D materials that are not gapped by spin-orbit couplings (SOCs) \cite{young2015dirac,Kowalczyk_2020,PtPb}, and thus exhibit behavior distinct from graphene \cite{PhysRevB.75.041401,kane2005quantum}, surface states of topological insulators \cite{RevModPhys.83.1057}, and 3D Dirac and Weyl semimetals \cite{young2012dirac,RevModPhys.90.015001}. 
They have more than one Dirac point in their first Brillouin zone (BZ). 
The band crossing at the Dirac points of these materials is protected by nonsymmorphic lattice symmetry \cite{young2015dirac}.
The 2D CuMnAs is a prototype of Dirac materials with a long-range AFM order \cite{tang2016dirac,CuMnAs1,maca2012room}. 
The recent discovery of electrical switching, via staggered spin-orbit torque, and readout of the N{\'e}el order vector in CuMnAs AFM layers, via anisotropic magnetoresistance \cite{wadley2016electrical,2ndAMR} and Voigt effect \cite{Saidl_2017},
was an important step towards next-generation AFM-based solid-state memory chips.
It was also shown that the fourfold degeneracy of Dirac points in this material can be lifted by manipulating nonsymmorphic symmetries. This is achieved by electrically altering the direction of the N{\'e}el vector through spin-orbit torques. Therefore a topological metal-insulator transition can happen in this system electrically \cite{vsmejkal2017electric}.

AC conductivity is a crucial parameter in condensed matter physics because it provides information on the electronic and magnetic behavior of materials in response to electromagnetic radiation, particularly in the optical frequency range \cite{Ebert_1996,Falkovsky_2007,graphene1,graphene2,PhysRevB.109.165441,PhysRevB.104.235403,Gusynin_2007,PhysRevB.94.014425,PhysRevLett.119.036601,rao2024tuneable,PhysRevLett.123.055901,PhysRevX.14.011052}. AC conductivity is related to various material properties such as dielectric functions, absorption spectra, charge susceptibilities, plasmon dispersion, and its linewidth, as well as magneto-optical effects \cite{mahan2013many,giuliani2008quantum,PhysRevB.109.045431,Ebert_1996}.
Very recently, terahertz time-domain spectroscopy was used to measure the anisotropic conductivity of CuMnAs thin films \cite{2023terahertz}.
Although there are a few studies on the optical properties of 2D Dirac materials in the presence of an external magnetic field with broken $\mathcal{PT}$ symmetry \cite{PhysRevB.105.085101,PhysRevB.107.245120}, to the best of our knowledge, no comprehensive investigation of the AC conductivity of 2D AFM CuMnAs layers that preserve $\mathcal{PT}$ symmetry \cite{2ndAMR} has been conducted so far \cite{cumnasDirac}. This article aims to bridge this gap by examining the AC conductivity, magnetic dichroism, and dielectric function of the 2D AFM CuMnAs layer. We specifically explore variations in these properties by tuning the orientation of the Néel vector by employing a time-dependent linear response Kubo formalism.

The rest of the paper is organized as follows. We first introduce the effective model Hamiltonian for a 2D AFM CuMnAs in Sec. \ref{model}. In Sec. \ref{AC}, we compute the AC conductivity of the system. Magnetic dichroism and dielectric function for different N{\'e}el vector orientations are investigated in Secs. \ref{MD} and \ref{Diel}, respectively. Finally, we summarize our results in Sec. \ref{summary}.

\section{Model}\label{model}
\begin{figure}[t]
\begin{center}
\includegraphics[scale=0.3]{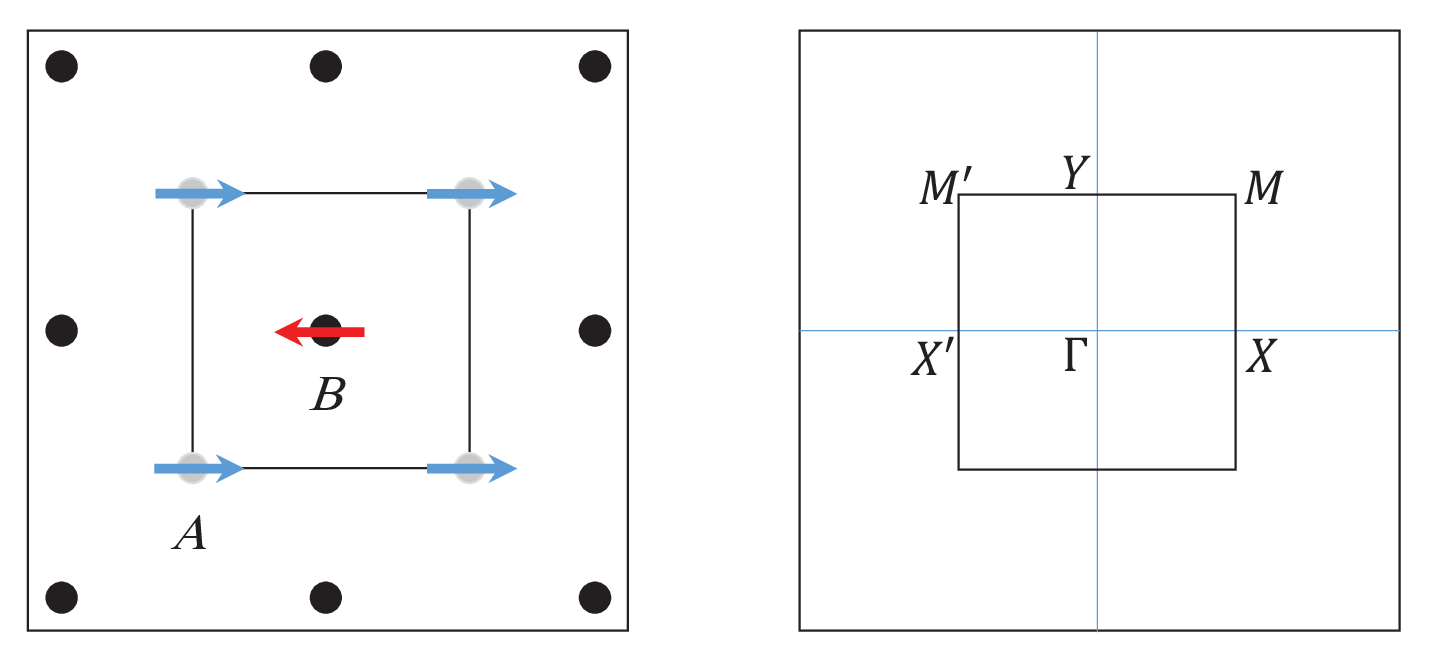}
\end{center}
\caption{Lattice structure for a 2D Dirac semimetal. $A$ (gray) and $B$ (black) sublattices are displaced in the out-of-plane $\hat{z}$ direction, which makes the lattice nonsymmorphic. This crinkling of the lattice permits a second neighbor spin-orbit interaction \cite{young2015dirac}. The blue and red arrows indicate the AFM magnetic moment directions on each sublattice. (Right) The first BZ and its high-symmetry points.}
\label{lattice}
\end{figure}

\begin{figure}[t]
\begin{center}
\includegraphics[scale=0.833]{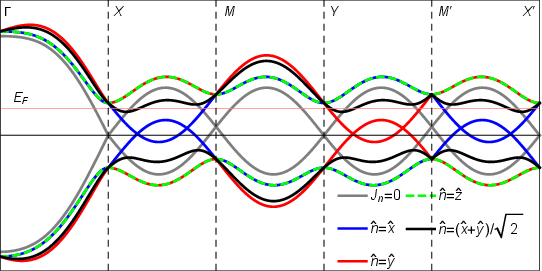}
\end{center}
\caption{
 The band structure of a 2D Dirac semimetal in the presence of AFM order for different directions of the N{\'e}el vector. The Fermi level, set at $E_F=0.5 t$, is shown with a horizontal pink line. In the absence of AFM order, $J_n=0$ (solid gray color line), there are three distinct Dirac nodes in the first BZ. The Dirac node at the M symmetry point is in different energy than the Dirac nodes at X and Y symmetry points. We set $t'=0.08t$, $\lambda=0.8 t$, and $J_{\rm n}=0.6 t$ \cite{vsmejkal2017electric}.}\label{dispersion}
\end{figure}

We employ the minimal model, introduced in Ref. \cite{vsmejkal2017electric}, to describe a tetragonal CuMnAs on a crinkled quasi-2D square lattice with a collinear AFM state, depicted in Fig. \ref{lattice}, where each atom is characterized by one orbital in the model Hamiltonian.
The total Hamiltonian consists of a kinetic term for conduction electrons $H_0$, a SOC term $H_{\rm{SO}}$, and an s--d(f) exchange interaction between the spin of itinerant electrons and localized AFM spins $H_{\rm{sd}}$,

\begin{subequations}\label{hamiltonian}
\begin{align}
  H  = &  H_0 + H_{\rm{SO}} + H_{\rm{sd}}, \\ 
  H_0  = &  -2 t \tau_x \sigma_0 \cos \frac{ak_x}{2} \cos \frac{ak_y}{2} \nonumber \\ &-  t'  \tau_0 \sigma_0 [\cos (ak_x)+\cos (ak_y)],\\
 H_{\rm{SO}}  = &  \lambda\tau_z[\sigma_y \sin (ak_x)-\sigma_x \sin (ak_y)],\\
 H_{\rm{sd}}  = &  J_{\rm{n}} \tau_z \bm{\sigma} \cdot \bm{n},
\end{align}
\end{subequations}
where $a$ is the lattice constant, $\bm{k}=\{k_x,k_y\}$ is the 2D electron wavevector, $\bm{\tau}$ and $\bm{\sigma}$ are $2 \times 2$ Pauli matrices describing the lattice and spin degrees of freedom, respectively; $\bm{n}$ is the N{\'e}el vector direction; $t$, $t'$, $\lambda$, and $J_{\rm{n}}$ parametrize the nearest-neighbor-hopping integral, next-nearest-neighbor hopping integral, next-nearest-neighbor SOC, and AFM s--d(f) exchange coupling strength, respectively.  
The Hamiltonian (\ref{hamiltonian}) has two doubly degenerate eigenvalues,
\begin{eqnarray}\label{eigenenergy}
E_{\bm{k}}^\pm=\gamma_2 \pm {\varepsilon_{\bm{k}}}, 
\end{eqnarray}
and the corresponding eigenvectors are given by
\begin{subequations}\label{eigenvectors}
\begin{eqnarray}
\psi_{1}^\pm&=&(-\gamma_1,\delta_z\mp \varepsilon_{\bm{k}},0,\delta_x+i\delta_y)^T/N_{\mp},\\
\psi_{2}^\pm&=&(\delta_z\pm \varepsilon_{\bm{k}},\gamma_1,\delta_x+i\delta_y,0)^T/N_{\pm},
\end{eqnarray}
\end{subequations}
where $T$ denotes the transpose symbol, 
$N_{\pm}=\sqrt{2 \varepsilon_{\bm{k}} \big(\varepsilon_{\bm{k}} \pm \delta_z\big)}$ is the normalization factor, and we define the following parameters:
\begin{subequations}
\begin{eqnarray}\label{params}
\gamma_1 &=& - 2 t \cos \frac{ak_x }{2} \cos \frac{ak_y }{2},\\
\gamma_2 &=& - t' [\cos (ak_x) +\cos (ak_y)],\\
\delta_x &=& J_{\rm{n}} n_x - \lambda \sin (ak_y) ,\\
\delta_y &=& J_{\rm{n}} n_y + \lambda \sin (ak_x) ,\\
\delta_z &=& J_{\rm{n}} n_z,\notag\\
\varepsilon_{\bm{k}} &=& \sqrt{\gamma_1^2+\delta_x^2+\delta_y^2+\delta_z^2}.
\end{eqnarray}
\end{subequations}
In the absence of AFM order, $J_{\rm{n}}=0$, there are three distinct Dirac points at the symmetry points $\bm{X}=\pm (\pi/a ,0)$, $\bm{Y}=\pm  (0,\pi/a )$, and $\bm{M}=\pm (\pi/a ,\pi/a )$ of the BZ, in which $\gamma_1=\delta_{x(y)}=0$. As we mentioned earlier, these Dirac points can even persist in the presence of SOC.
Figure \ref{dispersion} shows the electronic band structure, Eq. (\ref{eigenenergy}), in the absence of an AFM order ($J_{\rm{n}}=0$) with a solid gray line. Three Dirac points are located at the symmetry points $\bm{X}$, $\bm{Y}$, and $\bm{M}$ of the BZ. In Ref. \cite{young2015dirac} it is discussed that one of these Dirac points can be lifted by deforming the lattice, but not all of them.

In the presence of AFM order, it is possible to break the fourfold degeneracy of Dirac points and open a gap and have a semimetal-insulator transition. 
The condition for band crossing, $E_{\bm{k}}^+=E_{\bm{k}}^-$, requires $\varepsilon_{\bm{k}}=0$ and thus $\gamma_1=\delta_x=\delta_y=\delta_z=0$. 
To preserve the fourfold degeneracy at the Dirac points, where $\gamma_1=0$, in the presence of AFM order, the out-of-plane component of the N{\'e}el vector must be zero $n_z = 0$. 
In addition, the system must satisfy either of the following two criteria: $(i)$ at $k_x=\pm \pi/a$, we should have $\sin (ak_y) =J_{\rm{n}}/ \lambda \leq 1$, that requires $n_x=1$ and $n_y=0$; or $(ii)$ at $k_y=\pm \pi/a$, we should have $\sin (ak_x)=J_{\rm{n}}/\lambda \leq 1$, that requires $n_x=0$ and $n_y=1$. Therefore, the only directions of the N{\'e}el vector in which the fourfold degeneracy is preserved at two out of three Dirac points, are inplane directions along the $x$ and $y$ directions if the exchange coupling is smaller than the SOC parameter, $J_{\rm{n}}/\lambda\leq1$. It is evident that by changing the direction of the N{\'e}el vector, one can turn on and off the Dirac gap in this system. In addition, the positions of the Dirac cones are shifted by changing the direction of the N{\'e}el vector, see Fig. \ref{dispersion}. When the Néel vector has either an out-of-plane or inplane component between the $x$ and $y$ directions, all Dirac points become gapped; see electronic dispersion with the green and black curves in the bottom panel of Fig. \ref{dispersion}. Conversely, if the Néel vector lies inplane along the \(x\) or \(y\) directions, only some of Dirac points become gapped in first BZ, as shown by the blue and red curves in Fig. \ref{dispersion}. With the material parameters used in Fig. \ref{dispersion}, the system becomes an insulator when the Néel vector is aligned along the \(z\) direction.

\section{AC conductivity}\label{AC}
The AC conductivity in the clean limit within the Kubo formalism is given by \cite{bruus2004many,Ebert_1996},
\begin{eqnarray}\label{Kubo}
  \sigma_{\alpha\beta}(\omega)=\frac{i\hbar}{A}\sum_{mn}\frac{f_m-f_n}{E_m-E_n}\frac{\langle n|j_{\alpha}|m\rangle \langle m|j_{\beta}|n\rangle}{(E_m-E_n)-\hbar(\omega+i\eta)}
\end{eqnarray}
where $j_{\alpha}=(-e/\hbar)\partial H/\partial k_{\alpha}$ is the current density operator in the direction $\alpha$, $-e<0$ denotes the electron charge, $|n\rangle$ is the eigenstate, Eq. (\ref{eigenvectors}), with $n(m)$ represent the collective quantum numbers, including wavenumber $k$ and band index $\pm$; $f_n$ is the equilibrium Fermi-Dirac distribution function of the state $n$, $A$ is the area of the two-dimensional lattice, and $\eta$ parametrizes the impurity scattering rate. The optical conductivity consists of intraband (Drude) and interband contributions. In the present paper, we compute the AC conductivity at zero temperature.

\begin{figure}[t]
\begin{center}
\includegraphics[scale=0.8]{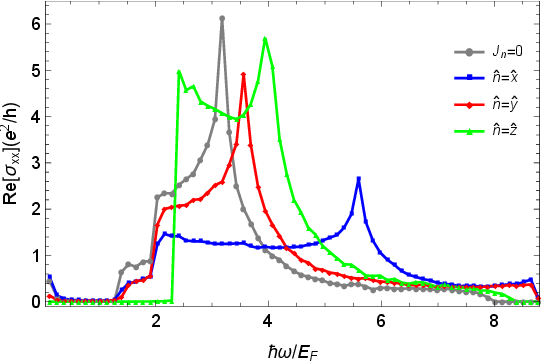}
\includegraphics[scale=0.8]{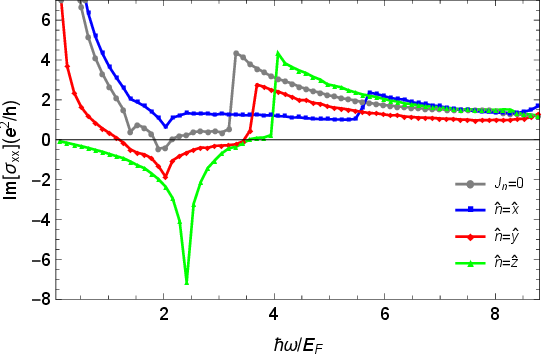}
\end{center}
\caption{The real (top) and imaginary (bottom) parts of the longitudinal conductivity along the $x$ direction in the absence (gray color line) and presence of the AFM order as a function of the frequency. The conductivity is changed by varying the N{\'e}el vector direction. There is no DC conductivity when the N{\'e}el vector is out-of-plane since the Fermi energy lies inside the electronic band gap and the system becomes an insulator. $\hbar\eta=0.002 E_F$}
\label{Rexx}
\end{figure}

The conductivity tensor satisfies the Onsager-Büttiker relation $ \sigma_{\alpha\beta}(\omega; \bm{n})=\sigma_{\beta\alpha}(\omega; -\bm{n})$. If the system exhibits \(\mathcal{PT}\) symmetry, the conductivity tensor contains only symmetric components, and there is no Hall effect. However, a transverse response and an anisotropic longitudinal response are still possible, depending on the relative direction of the applied electric field and the Néel vector in the presence of SOC. In the absence of AFM order or in the presence of an out-of-plane N{\'e}el vector, the off-diagonal elements of the conductivity tensor vanish and the longitudinal conductivity is isotropic. In general, if the N{\'e}el vector resides in the $xz$ or $yz$ planes, the transverse conductivity is zero. When the N{\'e}el vector has an inplane component, the longitudinal conductivity is anisotropic.

To better understand direct optical transitions around three Dirac points in the absence of the AFM order, we first linearize the dispersion around these three symmetry points.
Around the $\bm{M}=(\pi,\pi)$ symmetry point, the low-energy gapless Dirac-like dispersion is isotropics and reads,
\begin{eqnarray}\label{DiracM}
  E_{\bm{M}}^\pm \simeq -2t'\pm \lambda a q,
\end{eqnarray}
where $\bm{q}=(k_x,k_y)-\bm{M}$ is the wavevector measured from the $\bm{M}$ symmetry point.
The dispersion relation is isotropic around this symmetry point and the nodal point is shifted by a constant $-2t'$.
The minimum photon energy, required for the direct interband transitions around the $\bm{M}$ symmetry point is given by
\begin{eqnarray}\label{}
 \hbar \omega = E_{q_{\rm F}}^+- E_{q_{\rm F}}^- = 2\lambda a q_{\rm F},
\end{eqnarray}
where $ q_{\rm F}=(E_{\rm F}+ 2 t')/\lambda$ is the Fermi wavenumber. 

On the other hand, the electronic dispersion  around the $\bm{X}=(\pi,0)$ symmetry point, is given by an anisotropic gapless Dirac-like dispersion relation 
\begin{eqnarray}\label{DiracX1}
  E_{\bm{X}}^\pm \simeq \pm \sqrt{(t^2+\lambda^2) (aq_x)^2 + \lambda^2(a q_y)^2},
\end{eqnarray}
where $\bm{q}=(k_x,k_y)-\bm{X}$. Around the $\bm{Y}=(0,\pi)$ symmetry point, we should interchange the labels $q_x \leftrightarrow q_y$ in Eq. (\ref{DiracM}). 
As we can see, the above dispersion relation is anisotropic but particle-hole symmetric. In the absence of $t$, this dispersion is reduced to the isotropic Dirac dispersion relation. We can simply see that the minimum photon energy for this case is $\hbar \omega = 2E_F$. 

Figure \ref{Rexx} presents the real (top) and imaginary (bottom) parts of the longitudinal AC conductivity of this system along the $x$ direction in the absence $J_{\rm n} =0$ and presence of the AFM order $J_{\rm n} \neq 0$ with different N{\'e}el vector direction.
As we discussed earlier, the longitudinal part, in general, can be anisotropic in this AFM system in the presence of SOC and thus $\sigma_{xx} \neq \sigma_{yy}$. This leads to a magnetic linear dichroism, which we will discuss in the next section. 
The real part of the longitudinal conductivity represents the inphase current, which induces resistive Joule heating. In contrast, the imaginary part corresponds to the $\pi/2$ out-of-phase current. 
The sign of the imaginary part of the longitudinal conductivity indicates inductive behavior ${\rm {Im}}[\sigma_{xx}]>0$ and capacitive behavior ${\rm {Im}}[\sigma_{xx}]<0$. The bottom panel of the Fig. \ref{Rexx} shows that changing the N{\'e}el vector orientation in this system, control the sign of ${\rm {Im}}[\sigma_{xx}]$.

In the low-frequency regime, when the Néel vector is out of the plane, the longitudinal DC conductivity drops to zero within our parameters. This occurs because the Fermi energy (\(E_F\)) is smaller than the band gap of the system, rendering it an insulator. Conversely, if \(E_F\) exceeds the band gap, the intraband contribution dominates at low frequencies, leading to standard DC Drude conductivity. At higher frequencies, the interband contribution becomes the predominant contribution in the conductivity.
\begin{figure}[t]
\includegraphics[scale=0.8]{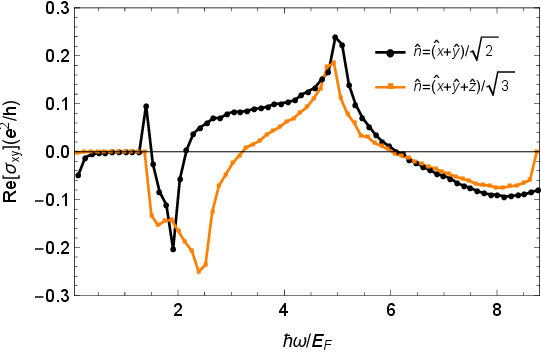}
\includegraphics[scale=0.8]{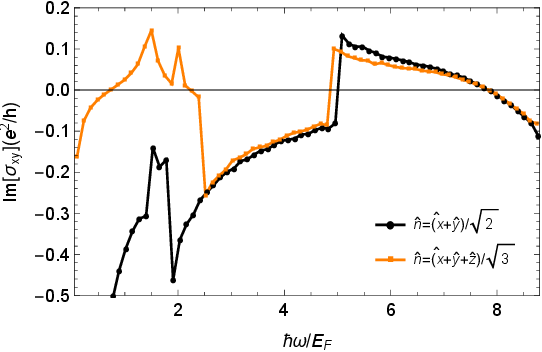}
\caption{The real (top) and imaginary (bottom) parts of the transverse conductivity in the presence of the AFM order as a function of the frequency. The sign and amplitude of the transverse conductivity are changed by varying the N{\'e}el vector direction. The transverse conductivity is zero when the N{\'e}el vector direction is in the $xz$ or $yz$ planes. We set $\hbar\eta=0.002 E_F$.}
\label{Rexy}
\end{figure}

The real and imaginary parts of the transverse AC conductivity of the system are plotted in the top and bottom panels of Fig. \ref{Rexy}, respectively. As we discussed earlier, this transverse conductivity is symmetric $\sigma_{xy}=\sigma_{yx}$ and thus dissipative, contrary to the Hall response, related to the antisymmetric part of the transverse response. When the Néel vector lies within the \(xz\) or \(yz\) planes, the system maintains mirror symmetry relative to these planes, resulting in zero transverse conductivity. Conversely, if the Néel vector has an inplane component between the \(x\) and \(y\) directions, this symmetry is broken, leading to a finite transverse conductivity. Figure \ref{Rexy} shows that both the sign and the amplitude of the real and imaginary parts of the transverse conductivity can be tuned by changing the orientation of the Néel vector.

In general, Figs. \ref{Rexx} and \ref{Rexy} show that the AC response of the system is very sensitive to the N{\'e}el vector orientation.

\section{Magnetic dichroism} \label{MD}
Magnetic circular and linear dichroism are optical techniques for measuring the electronic band structure and the magnetic orientation of the system \cite{Ebert_1996}. These two effects are related to different components of the AC conductivity tensor.  
Circular dichroism is the differential absorption of left and right circularly polarized light in a magnetically ordered material. This effect is finite in chiral systems in which antisymmetric part of the off-diagonal elements of the AC conductivity tensor is finite \cite{PhysRevB.107.245120,Catarina_2020, PhysRevB.105.014437,Gudelli_2019,PhysRevB.101.045426}. In the Dirac AFM system, considered in this study, because of the $\mathcal{PT}$ symmetry, there is no antisymmetric (anomalous Hall) component \cite{2ndAMR} and therefore no circular dichroism. Therefore, both magneto-optical Faraday and Kerr effects are absent in this system. This fact makes the optical readout of the staggered order parameter challenging. 
\begin{figure}[t]
\begin{center}
\includegraphics[scale=0.8]{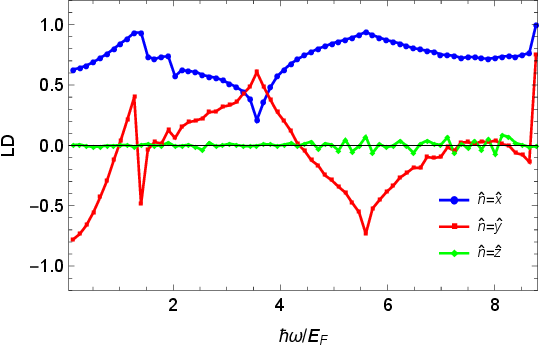}
\end{center}
\caption{The magnetic linear dichroism (LD) as a function of the frequency, Eq. (\ref{LDequation}). The amplitude and sign of the LD signal are sensitive to the direction of the N{\'e}el vector and frequency.}
\label{LD}
\end{figure}

On the other hand, the magnetic linear dichroism is the difference between the absorption of light polarized parallel and that polarized perpendicular to an orientation axis. This effect is related to the anisotropy in the longitudinal components of the AC conductivity tensor.
Although the Faraday and Kerr effects are first order in relation to the local magnetic moment, linear dichroism may arise from the Voigt-Cotton–Mouton effect \cite{Ferre_1984}, which is proportional to the square of the local magnetic moment. Additionally, it may stem from electronic anisotropy, associated with symmetry breaking because of changes in the direction of the Néel vector \cite{LD1,LD2}.
Hence, linear dichroism is a powerful tool for directly exploring symmetry breaking in the system.
The linear dichroism is defined as
\begin{eqnarray}\label{LDequation}
 {\rm LD}(\omega)=\frac{{\rm{Re}}[\sigma_{xx}] - {\rm{Re}} [\sigma_{yy}]}{{\rm{Re}} [\sigma_{xx}] + {\rm{Re}} [\sigma_{yy}]}.
\end{eqnarray}
We plot the linear dichroism in Fig. \ref{LD} for different N{\'e}el vector orientations. It is evident from this plot that switching the direction of the N{\'e}el vector, from out-of-plane to inplane, enhances the linear dichroism signal. The sign and amplitude of the signal can be changed by the light frequency and N{\'e}el vector direction. The sign of LD indicates the preferential absorption of light polarized in specific directions within an anisotropic material. This information is crucial for understanding the material's anisotropic properties.

\begin{figure}[t]
\begin{center}
\includegraphics[scale=0.8]{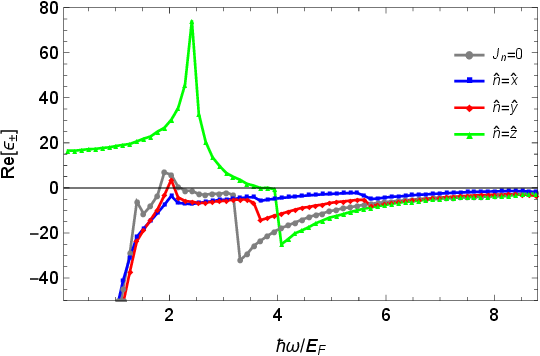}
\includegraphics[scale=0.8]{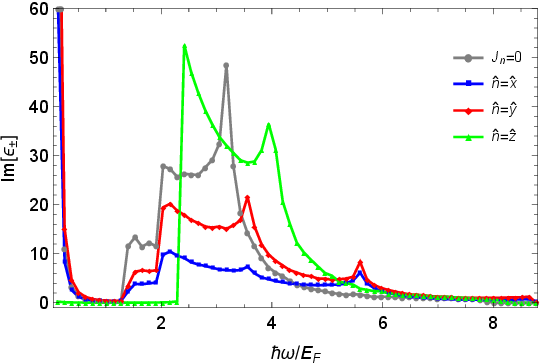}
\end{center}
\caption{Real (top panel) and imaginary (bottom panel) parts of the dynamic dielectric function, Eq. (\ref{Dielectric}), at long-wavelength limit $(k= 4.4 \times 10^{-11} {\rm m}^{-1})$, as a function of the frequency in the absence (gray line) and presence of the AFM order for various N{\'e}el vector directions. We set $\epsilon_0= 2.5 \kappa$, where $\kappa$ is the vacuum permittivity.}
\label{dielectric}
\end{figure}

\section{Dielectric function and reflective index}\label{Diel}
The dynamical dielectric permittivity $\epsilon_{\pm}(\omega)$ quantifies a substance's ability to hold an electrical charge, many-body charge screening, and in addition, encodes information about collective plasmon excitations, $\rm{Re}[\epsilon_{\pm}]=0$, in the system. This quantity is also related to the refractive index $n_{\pm}=\sqrt{\epsilon_{\pm}}$ \cite{Ebert_1996}, which quantifies the bending of electromagnetic waves in the medium. In 2D systems, the dynamical dielectric function, within the random phase approximation at the long-wavelength limit, reads \cite{bruus2004many}
\begin{equation} \label{Dielectric}
        \lim_{k \to 0} \epsilon_{\pm}(\omega,k) \approx 1 + i \frac{2\pi k }{\epsilon_0 \omega}\sigma_{\pm}(\omega),
    \end{equation}
where $\epsilon_0$ is the static permittivity of the substrate and $\sigma_\pm=\big[(\sigma_{xx}+\sigma_{yy}) \pm i(\sigma_{xy}-\sigma_{yx})\big]/2$ is the AC conductivity for the left (-) and right (+) circularly polarized light.
Since there is no antisymmetric contribution in the transverse conductivity tensor, the dielectric function is the same for both polarizations $\epsilon_+=\epsilon_-$, and thus $n_+=n_-$. Therefore there is no circular birefringence in the system under study. However, since in general $n_{xx} \neq n_{yy}$ in this system, there might be a finite linear birefringence \cite{Ferre_1984}, which can be tuned by the N{\'e}el vector orientation. 

Figure \ref{dielectric}, represents the real (top panel) and imaginary (bottom panel) parts of the dielectric function and shows that the tuning of the N{\'e}el vector changes the dielectric response and therefore the optical properties of the system.

\section{Summary and Concluding remarks}\label{summary}
In this study, we have investigated the effect of the N{é}el vector orientation on the dynamical response of CuMnAs, a prototype of a 2D Dirac antiferromagnetic semimetal with $\mathcal{PT}$ symmetry. 
We computed the AC conductivity, linear dichroism, and dielectric function of the system. Our calculations show that the orientation of the magnetic state may alter the system's optical response drastically. 
Because of the presence of \(\mathcal{PT}\) symmetry in our model, there is no anomalous Hall response. This distinguishes our AFM model from a nonsymmorphic 2D Dirac semimetal in the presence of an out-of-plane magnetic field, which exhibits a finite anomalous Hall response
\cite{PhysRevB.107.245120}. 
We propose that magnetic linear dichroism and birefringence are powerful tools for detecting the magnetic states in these systems.

\section*{Acknowledgements}
R.S. and S.E. would like to acknowledge financial support from the research council of the University of Tehran.
A.Q. acknowledges financial support from the Research Council of Norway through its Centers of Excellence funding scheme, project number 262633, ``QuSpin".

\bibliography{ref}

\end{document}